\documentclass[aps,prb,reprint,twocolumn,superscriptaddress]{revtex4-2}
\usepackage{graphicx}
\usepackage{amsmath}
\usepackage{amsfonts}
\usepackage{color}
\usepackage{amssymb}%
\usepackage{siunitx}
\usepackage{blindtext}

\usepackage{xspace}
\usepackage{siunitx} 
\usepackage{miller}

\usepackage[version=3]{mhchem}
\usepackage{natbib}
\usepackage[]{hyperref}

\usepackage{ulem}
\usepackage{bm}

\renewcommand{\vec}[1]{{\boldsymbol #1}}
\newcommand{\tbmno}{TbMnO$_3$ }

\newcommand{\mnwo}{MnWO$_4$ }

\newcommand{\nafegeo}{NaFeGe$_2$O$_6$ }
\newcommand{\nafegeokomma}{NaFeGe$_2$O$_6$}

\newcommand{\dmi}{Dzyaloshinskii-Moriya interaction }

\begin{document}


\title{Chiral Order and Multiferroic Domain Relaxation in \nafegeo}

\author{S. Biesenkamp}\email[e-mail: ]{biesenkamp@ph2.uni-koeln.de}
\affiliation{$I\hspace{-.1em}I$. Physikalisches Institut,
Universit\"at zu K\"oln, Z\"ulpicher Straße 77, D-50937 K\"oln,
Germany}

\author{D. Gorkov}
\affiliation{$I\hspace{-.1em}I$. Physikalisches Institut,
Universit\"at zu K\"oln, Z\"ulpicher Straße 77, D-50937 K\"oln,
Germany}

\author{W. Schmidt}
\affiliation{Juelich Centre for Neutron Science JCNS, Forschungszentrum Juelich GmbH, Outstation at ILL, 38042 Grenoble, France}

\author{K. Schmalzl}
\affiliation{Juelich Centre for Neutron Science JCNS, Forschungszentrum Juelich GmbH, Outstation at ILL, 38042 Grenoble, France}

\author{Y. Sidis}
\affiliation{Universit\'eParis-Saclay, CNRS, CEA, Laboratoire L\'eon Brillouin, F-91191, Gif-sur-Yvette, France}

\author{P. Becker}
\affiliation{Abteilung  Kristallographie,  Institut  f\"ur  Geologie  und  Mineralogie, Universit\"at zu K\"oln, Z\"ulpicher Straße  49b,  50674  K\"oln,  Germany}

\author{L. Bohat\'{y}}
\affiliation{Abteilung  Kristallographie,  Institut  f\"ur  Geologie  und  Mineralogie, Universit\"at zu K\"oln, Z\"ulpicher Straße  49b,  50674  K\"oln,  Germany}

\author{M. Braden}\email[e-mail: ]{braden@ph2.uni-koeln.de}
\affiliation{$I\hspace{-.1em}I$. Physikalisches Institut,
Universit\"at zu K\"oln, Z\"ulpicher Straße 77, D-50937 K\"oln,
Germany}





\date{\today}

\begin{abstract}
The magnetic structure and the multiferroic relaxation dynamics of NaFeGe$_2$O$_6$ were studied by neutron scattering on single crystals partially utilizing polarization analysis. In addition to the previously reported transitions, the incommensurate spiral ordering of Fe$^{3+}$ moments in the $ac$ plane develops an additional component along the crystallographic $b$ direction below $T\approx\SI{5}{\kelvin}$, which coincides with a lock-in of the incommensurate modulation. The quasistatic control of the spin-spiral handedness, respectively of the vector chirality, by external electric fields proves the invertibility of multiferroic domains down to the lowest temperature. Time-resolved measurements of the multiferroic domain inversion in NaFeGe$_2$O$_6$
reveal a simple temperature and electric-field dependence of the multiferroic relaxation that is well described by a combined
Arrhenius-Merz relation, as it has been observed for TbMnO$_3$. The maximum speed of domain wall motion is comparable to the spin wave velocity.

\end{abstract}


\maketitle


\section{\label{sec:level1}Introduction}

The increasing power consumption due to data storage and the growing amount of information that has to be stored or buffered enforce the development of new and more effectively working memory devices \cite{jones2018,Puebla_2020}. Beside, phase change based devices, skyrmion racetracks and other promising techniques \cite{Ceze_2019,Lin_2020,Wuttig_2007,Tomasello_2014,Puebla_2020,Tizno_2019,Ghosh_2016,Gu_2014}, multiferroic systems are promising for designing future storage techniques \cite{Scott2007a,Spaldin2019}.

In multiferroics at least two ferroic ordering parameters occur in the same phase and are coupled to each other \cite{Khomskii2009}. Both ordering parameters can be controlled by both conjugate fields. The magnetoelectric case with ferroelectric and magnetic ordering is most prominent, as it allows for a multiferroic memory, which combines the advantages of ferroelectric (FeRAMs) and magnetic random access memories (MRAMs) \cite{Scott2007a,Spaldin2019,Fiebig2016}. In so-called type-II multiferroics the ferroelectric polarization is not only coexisting with magnetic ordering but it is also induced by the onset of it, which hence implies strong coupling between both ferroic ordering parameters \cite{Khomskii2009}. In many systems the driving mechanism for this effect is the inverse \dmi (DMI), for which a spin canting of neighboring spins $\vec{S_i}$ and $\vec{S_j}$ induces a ferroelectric polarization, whose direction is defined by $\vec{P}\propto\vec{e_{ij}}\times\left(\vec{S_i}\times\vec{S_i}\right)$ with $\vec{e_{ij}}$ being the connecting vector of neighboring spins \cite{Dzyaloshinsky1958,Moriya1956,Mostovoy2006}.

The understanding of the read and write performance under different conditions is indispensable for the development of new memory devices. The control and the dynamics of multiferroic domain inversion were studied intensively for \tbmno and \mnwo using neutron scattering techniques, second-harmonic generation (SHG) measurements and dielectric spectroscopy \cite{Stein2017,Baum2014,Hoffmann2011a,stein2020,Niermann_2014}. The relaxation behavior in both systems differs significantly as a function of temperature and electric-field amplitude \cite{Baum2014,Hoffmann2011a,stein2020}. In \mnwo domain inversion speeds up, when cooling towards the lower commensurate magnetic phase \cite{Baum2014,Hoffmann2011a}, which is attributed to depinning of multiferroic domains by anharmonic modulations \cite{Finger2010,biesenkamp2020}.
In \tbmno the domain inversion can be well described by a remarkably simple law combining an Arrhenius-like activated temperature dependence with a
field dependence following the Merz law established for ferroelectrics \cite{stein2020}.
In contrast to MnWO$_4$, the incommensurate long-range order in multiferroic \tbmno does not transform into a commensurate arrangement \cite{Kajimoto2004,Kimura2005} but
is only altered through the additional order of Tb$^{3+}$ moments \cite{Prokhnenko_2007}.
Apparently, competing incommensurate and commensurate ordering significantly influence the relaxation behavior of multiferroic domain inversion and thus complicate its description \cite{biesenkamp2020}. Therefore, multiferroic materials with simple phase diagrams are desirable for further investigations of multiferroic domain dynamics.

The broad class of pyroxenes with chemical formula $A\text{\textit{Me}}X_2$O$_6$ with $A$ = Li, Na, Sr or Ca, $X$ = Si or Ge and \textit{Me} a magnetic transition metal exhibit several interesting
magnetic ordering phenomena, with controllable ferrotoroidicity in LiFeSi$_2$O$_6$ \cite{Jodlauk2007,Baum2013} and multiferroic phases in NaFeSi$_2$O$_6$ \cite{Jodlauk2007}, in
\nafegeo \cite{Kim_2012} and in  SrMnGe$_2$O$_6$ \cite{Ding2016} being the most prominent ones. In spite of the similar monoclinic crystal structure with magnetic zigzag chains running along the $c$ direction, see Fig. 1, the multiferroic mechanism is different in these three pyroxenes. In NaFeSi$_2$O$_6$ two successive transitions involving the same
irreducible representation result in a helical structure with the propagation vector along the monoclinic axis \cite{Baum2015}. In this material multiferroic order does not
result from the spin-current or inverse Dzyaloshinski-Moriya mechanism \cite{Dzyaloshinsky1958,Moriya1956,Katsura2005,Mostovoy2006} but from the combination of the chiral order and the monoclinic distortion. In contrast, multiferroic order in SrMnGe$_2$O$_6$ \cite{Ding2016} and in \nafegeo \cite{Ding_2018} seems to follow the most common inverse Dzyaloshinski-Moriya mechanism \cite{Dzyaloshinsky1958,Moriya1956,Katsura2005,Mostovoy2006} but with
differently oriented cycloidal and ferroelectric order.

\begin{figure}
 \includegraphics[width=\columnwidth]{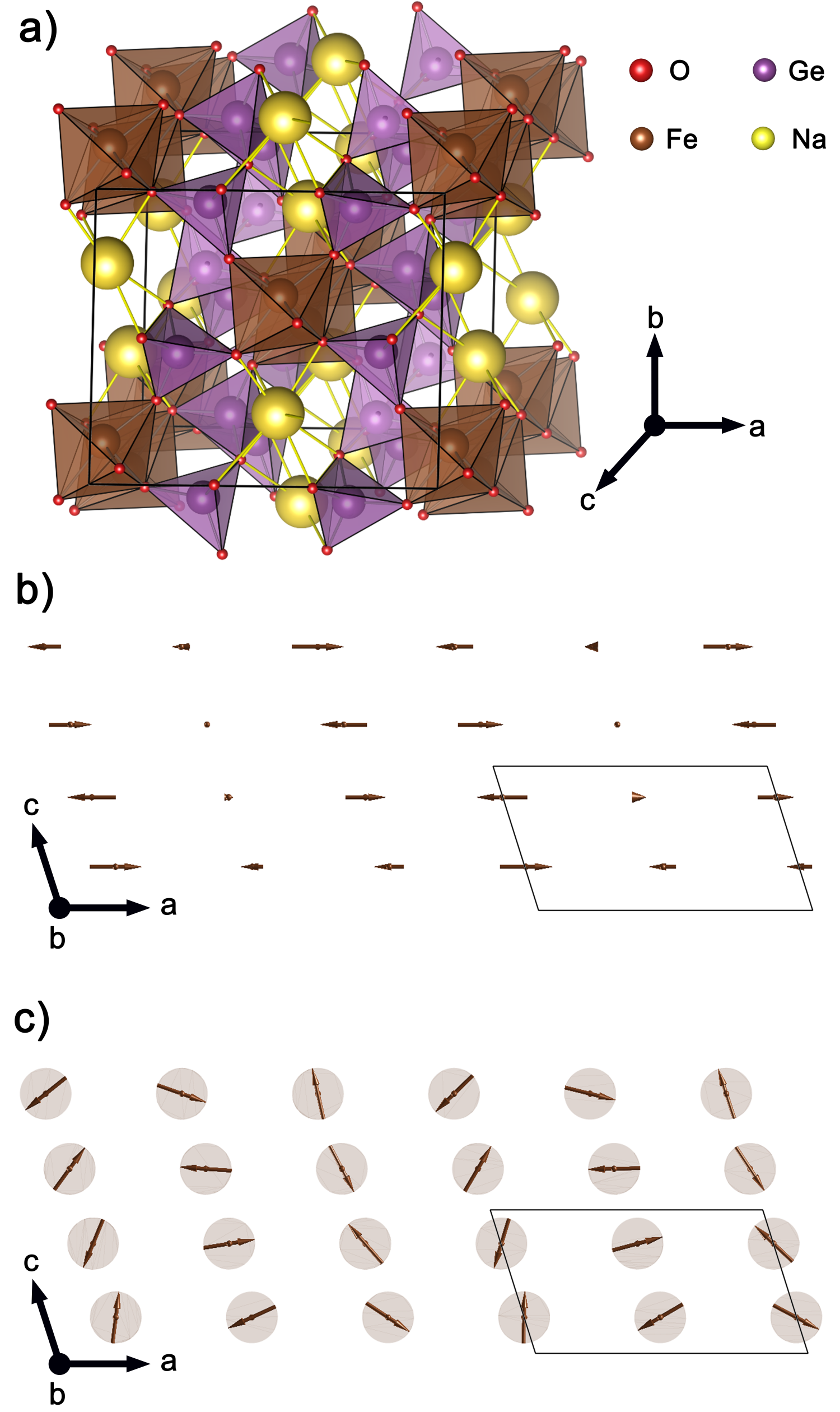}
  \caption{\label{fig:structure}The nuclear structure (with data from Ref. \onlinecite{Redhammer_2011}) is shown in panel a) by utilizing the crystallographic visualization software \textsc{VESTA3} \cite{Momma2011}. Both panels b) and c) sketch the magnetic structure of the intermediate SDW phase and of the low temperature spin-spiral phase, respectively.}
  \label{disp}
 \end{figure}

\nafegeo exhibits a simple phase diagram with the typical sequence of two incommensurate ordered phases and with just one magnetic site  \cite{Kim_2012,Drokina_2011,Redhammer_2011,Ackermann_2015,Ding_2018}. The system crystallizes in the monoclinic spacegroup $C2/c$ with $a=\SI{10.0073(8)}{\angstrom}$, $b=\SI{8.9382(7)}{\angstrom}$, $c=\SI{5.5184(4)}{\angstrom}$ and $\beta=\SI{107.524(1)}{\degree}$ \cite{solovyova1967x,Redhammer_2011}. Edge sharing FeO$_6$ octahedra form zigzag chains along the $c$ direction and in $b$ direction, these chains are separated by corner sharing GeO$_4$ tetrahedra that are also stacked along $c$ (see Fig. \ref{fig:structure}).
Below $T\approx\SI{35}{\kelvin}$, short-range ordering was observed and two different magnetic phases with incommensurate long-range order were reported for \nafegeo with transition temperatures at $T_\text{N}\approx\SI{13}{\kelvin}$ and $T_\text{MF}\approx\SI{11.6}{\kelvin}$ \cite{Drokina_2011,Redhammer_2011,Ackermann_2015,Ding_2018}. First at $T_\text{N}$ an incommensurate spin-density wave (SDW) evolves with moments pointing roughly along $a$-direction \cite{Ding_2018} and below $T_\text{MF}$, spins form a chiral spin structure with moments lying within the $ac$ plane \cite{Drokina_2011,Redhammer_2011}. A small $b$ component of the chiral structure was controversially  discussed \cite{Drokina_2011,Redhammer_2011}, and in \cite{Redhammer_2011} only a single transition is observed. Simultaneous to the onset of the chiral spin arrangement, a ferroelectric polarization ($\vec{P}\approx\SI{32}{\micro\coulomb\per\meter\squared}$) develops \cite{Ackermann_2015,Kim_2012}. The ferroelectric polarization is confined to the $ac$-plane as the given symmetry forbids a non-zero component of the polarization along $b$ \cite{Ding_2018}, see section III, but symmetry allows for a finite $b$-component of the chiral spin structure.

The simplicity of the phase diagram makes \nafegeo an ideal candidate to investigate multiferroic domain dynamics. In the following sections we will first discuss the presence of the magnetic $b$ component utilizing neutron polarization analysis. Subsequently we will report on our investigations of multiferroic domain inversion in \nafegeokomma{.} Finally we will discuss the measured spin-wave velocity, which is proposed to limit the inversion speed of multiferroic domains.

 \section{\label{sec:experimental_methods}Experimental Methods}
A detailed description of the \nafegeo single crystal growth by the top seeding technique can be found in Ref. \onlinecite{Ackermann_2015}. For our measurements two single crystals (SI and SII)  were prepared and had been characterized by susceptibility measurements utilizing a commercial superconducting quantum interference device (SQUID) magnetometer. For both samples the transition temperatures are slightly reduced to $T_\text{N}\approx\SI{12.5}{\kelvin}$ and $T_\text{MF}\approx\SI{11.1}{\kelvin}$ with respect to literature \cite{Ackermann_2015}. In order to apply electric fields to the system, both samples were mounted between aluminium plates, which can be connected to a high voltage setup. As the ferroelectric polarization is largest along the $a$ direction \cite{Ackermann_2015}, the plate normal was oriented parallel to it. The sample surface was thinly coated with silver paste for optimized electric contact and the aluminum plates were tightened together by insulating polytetrafluorethylene (PTFE) screws (see Ref. \onlinecite{stein2020}). The plates that are cramping the respective sample were connected to sample holders defining the scattering geometry (1 0 0)/(0 0 1) for sample SI and (1 0 0)/(0 1 0) for sample SII.

Neutron polarization analysis senses the different components of the magnetic structure and the sign of the vector chirality. The respective experiments were performed at the cold three-axes spectrometer IN12 (located at the Institut Laue-Langevin) and at the cold three-axes spectrometer 4F1 (located at the Laboratoire L\'eon Brillouin).

At IN12, a horizontally and vertically focussing pyrolytic graphite (002) monochromator supplied $\lambda=\SI{3}{\angstrom}$ and a supermirror cavity provided a highly polarized neutron beam (flipping ratio of about FR$\approx$24). Higher order contaminations were suppressed by a velocity selector. The IN12 spectrometer was also equipped with a Helmholtz-coil setup in order to define the guide-field direction at the sample position, and spin-flippers before and after the sample position enabled longitudinal polarization analysis. A curved Heusler (111) analyzer selected the final beam polarization.
A conventional high-voltage generator (\textsc{Fug HCP 14-3500}) was deployed for measurements with quasistatic electric fields ($|U|\leq\SI{3.5}{\kilo\volt}$)
but a stroboscopic high-voltage setup was used for the time-resolved investigation of the multiferroic domain relaxation. This setup contains two high-voltage modules (\textsc{Iseg BPP4W} and \textsc{Iseg} BPN4W) providing high voltage ($|U|\leq\SI{4}{\kilo\volt}$) with opposite polarity and a fast \textsc{MOSFET} array (\textsc{Behlke HTS-111}), which is capable to switch the high voltage output polarity with $\SI{50}{\micro\second}$ \cite{stein2020}.  A multichannel data collector (\textsc{Mesytec MCPD-8}) records all detector events as well as a synchronization signal, when the polarity of the applied electric-field is switched. All events are recorded together with a timestamp, which allows to periodically switch the field, while summing the detected neutrons in time bins with respect to the moment when the field is reversed.
This stroboscopic method thus enables one to investigate very fast relaxation processes, because the width of each time bin can be arbitrarily selected, while repeating the switching periodically yields the required statistics.
The time resolution is limited by the finite spreads of neutron velocities and path lengths.
Ref. \onlinecite{stein2020} gives a detailed description of this setup and of the stroboscopic technique.

For long relaxation times above several minutes, the counting statistics obtained during a single switching period is sufficient and does not require the stroboscopic setup. The respective measurements of long relaxation times were performed at the cold three-axes spectrometer 4F1. A double pyrolytic graphite monochromator set-up supplied $\lambda=\SI{4.2}{\angstrom}$ and a supermirror bender polarized the neutron beam (flipping ratio of about $\text{FR}\approx50$). For polarization analysis a MuPad system \cite{Janoschek2007} and a horizontally curved Heusler analyzer were installed.
In order to measure the spin-wave velocity, a polarized beam is not required and the respective experiment was conducted at the 4F2 spectrometer. The double pyrolytic-graphite monochromator provided an unpolarized neutron beam with $\lambda=\SI{4.05}{\angstrom}$ and the pyrolytic-graphite analyzer enables inelastic measurements, for which a cooled Be filter was additionally inserted.

 \section{\label{sec:magnetic_strcuture}Magnetic structure}

The magnetic structure of \nafegeo was not fully characterized, as a $b$ component of the chiral spin structure was controversially discussed \cite{Drokina_2011,Redhammer_2011} and as the determination of the intermediate structure is based on neutron powder diffraction experiments \cite{Drokina_2011,Ding_2018}. Neutron polarization analysis on single crystals permits a precise separation of magnetic components and unambiguous identification of chiral signatures.

The symmetry conditions of incommensurate magnetic order with a propagation vector $\vec{k}=(k_h\:0\:k_l)$
have been discussed for \nafegeo in reference \cite{Ding_2018}; they are furthermore identical to
those in multiferroic LiFe(WO$_4$)$_2$ \cite{Liu2017,Biesenkamp2021}. In space group $C2/c$ only one symmetry element has to be considered, the glide mirror plane $c$, which transforms $(m_x,m_y,m_z)$ into $(-m_x,m_y,-m_z)$. Therefore, it is obvious that magnetic $x,z$ and $y$ components are separated.
The character concerning this $c$ element is plus or minus 1 up to the phase correction.
The inverse of this phase enters the basis vector components for the second site.
In the magnetic superspace group formalism all magnetic moments are developed in cosine and sine series starting from the atoms in the
primitive unit cell. Here we have two magnetic ions in the unit cell and the coefficients of these two are not independent but
determined through the condition that the moment at $-{\bf r}$ and $-x_4$ (the argument of the modulation functions) is identical
to the moment at  ${\bf r}$ and $x_4$, which is just the inversion symmetry \cite{Perez-Mato2012}.
Therefore, only one set of cosine and sine functions is required in the harmonic description.
For one-dimensional irreducible representations, as it is the case here, there is a one-to-one correspondence
between irreducible representation and magnetic superspace group formalisms, but the representation analysis leaves the phases undefined, while magnetic superspace group analysis
fixes these phases leaving only three free parameters for each of the magnetic modes \cite{Perez-Mato2012}.
The symmetry of the two different incommensurate magnetic modes is given in Table I.

The magnetic superspace group analysis explains that a multiferroic cycloid cannot be obtained with a single
symmetry mode, as the inversion symmetry is not broken \cite{Perez-Mato2012}. One may also argue that in each of the two modes the $x$ and $z$ components are modulated in
phase, while  $x,z$ and $y$ components have a different character between the two sites so that any spin-current contributions
cancel out, as it is also the case in the intermediate phase in MnWO$_4$ \cite{Urcelay-Olabarria2013}.
In order to generate cycloidal order one needs to couple two magnetic modes with a phase shift of $(\frac{\pi}{2}+n\pi)$ by either combining the two
different magnetic modes or by combining twice the same mode with the phase shift. In the former case the glide-mirror symmetry is lost and
ferroelectric order appears along the $b$ direction as in LiFe(WO$_4$)$_2$ \cite{Liu2017,Biesenkamp2021}, while in the latter case the mirror plane
persists and enforces ferroelectric polarization in the $a,c$ plane \cite{Ding_2018}. This is the situation in the multiferroic phase of
\nafegeokomma{.}

\begin{table*}
\caption{\label{tab:character_table}  Symmetry conditions for the two incommensurate magnetic modes that can appear for a propagation vector $\vec{k}=(k_h\:0\:k_l)$
in space group $C2/c$ with a single magnetic ion at (0,0.904,0.25). The conditions are defined by the irreducible representation $\Gamma_1$ and $\Gamma_2$ and the corresponding basis vectors for the two Fe sites are given  with  $a = e^{i2\pi{\bf \delta r}{\bf q_{inc}}}$. Here ${\bf \delta r}={\bf r}_2-{\bf r}_1$ and $a^*=1/a$ is the complex conjugate of $a$. Representation analysis and superspace group formalism
are equivalent in the case of \nafegeokomma{,} but the superspace group formalism fixes the phases between modulations so that for each of the
two modes the magnetic structure is described by only three parameters corresponding to either cosine or sine modulations. The resulting superspace groups
for a single mode are given in the last column.
}
\begin{ruledtabular}
\begin{tabular}{c|cc|cc|ccc}

$\Gamma$  &  1 & $c$ & ($x$,$y$,$z$) &($x$,$\bar{y}$,$z+1/2$) & coefficients         & character    & superspace group  \\ \hline
$\Gamma_1$&1 & $-a$  & ($u$,$v$,$w$) &$a^*$($u$,$-v$,$w$)     & $\text{Re}(u) \text{Im}(v) \text{Re}(w)$  & cos/sin/cos  &     $C2/c1'(\alpha 0 \gamma)0ss$          \\
$\Gamma_2$&1 & $a$   & ($u$,$v$,$w$) &$a^*$($-u$,$v$,$-w$)    & $\text{Im}(u) \text{Re}(v) \text{Im}(w)$  & sin/cos/sin  &  $C2/c1'(\alpha 0 \gamma)00s$        \\

\end{tabular}
\end{ruledtabular}
\end{table*}

For neutron polarization analysis it is convenient to define a right-handed coordinate system, where  the $x$ direction is set parallel to the scattering vector $\vec{Q}$, $y$ is perpendicular to $x$ but confined to the scattering plane and $z$ is perpendicular to both, $x$ and $y$. Magnetic scattering will only arise from magnetic components perpendicular to the scattering vector and spin-flip (SF) processes necessitate a component of the magnetization perpendicular to the direction of the incoming beam polarization. In contrast, non-spin-flip (NSF) processes only sense components that are aligned parallel or antiparallel to the incoming neutron beam polarization.
\begin{figure}
 \includegraphics[width=\columnwidth]{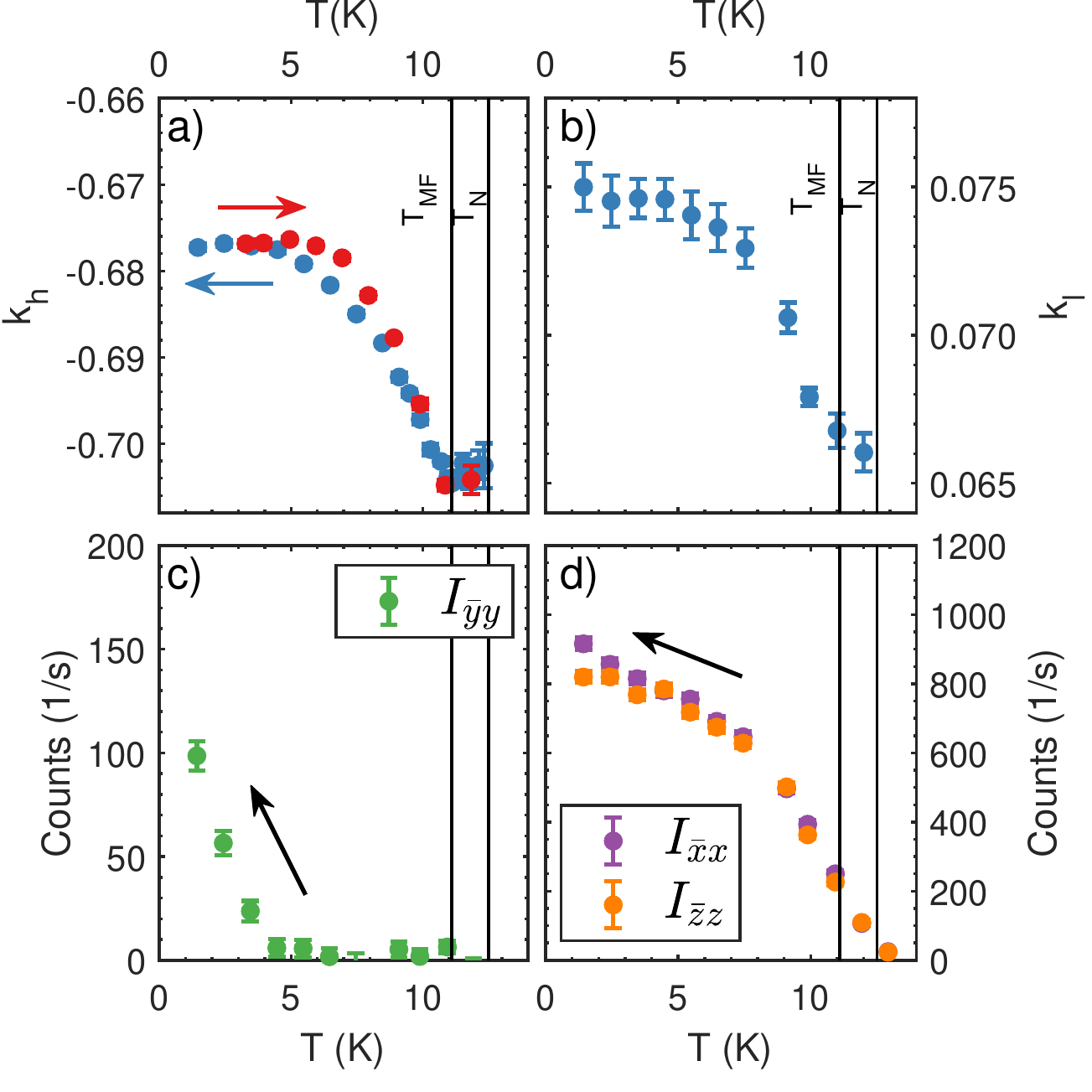}
  \caption{\label{fig:k_vector}Panels a) and b) display the temperature dependent components $k_h$ and $k_l$ of the propagation vector $\vec{k}=(k_h\:\:0\:\:k_l)$, respectively. The blue and red data correspond to cooling and heating sequences, respectively. In c) and d) the peak intensity of reflection $\vec{Q}=(0.678\:\:0\:\:\text{-}1.075)$ is plotted for all SF channels as function of temperature. All intensities were corrected for the finite flipping ratio and all measurements were performed on sample SI.}
  \label{disp}
 \end{figure}
During the first part of the allocated beamtime at IN12, sample SI with scattering geometry (1 0 0)/(0 0 1) was mounted. Both components of the incommensurate propagation vector $\vec{k}=(k_h\:\:0\:\:k_l)$ could be traced as a function of temperature. For the respective measurements, $Q_h$ and $Q_l$ scans were executed across the magnetic reflection $\vec{Q}=(0.678\:\:0\:\:\text{-}1.075)$ and each $Q$ scan was fitted by a Gaussian function. The observed peak center values were considered for defining the scan  of the following step in temperature.  Fig. \ref{fig:k_vector} a) and b) display the obtained temperature dependent values for $k_h$ and $k_l$, and it can be seen that both components vary with temperature but lock in a constant value at $T\approx\SI{5}{\kelvin}$. The results qualitatively agree with the reported temperature dependence of the incommensurate propagation vector in Ref. \onlinecite{Redhammer_2011}.

The used scattering geometry entails, that the SF channels are described by  $I_{\bar{y}y}\propto M_{b}M_{b}^*$ and $I_{\bar{z}z}\propto M_{ac}M_{ac}^*\text{sin}^2(\alpha)$ with $M_{b}$ and $M_{ac}$ being the complex Fourier components of the magnetization along $b$ direction and within the $ac$ plane, respectively, and $\text{sin}^2(\alpha)$ being the geometry factor. $a$ and $a^*$ are not parallel in the monoclinic system, but in the used scattering geometry both are lying within the scattering plane. The same holds for $c$ and $c^*$. Due to the incommensurability of the magnetic structure, no nuclear signal contributes to the measured intensity. The SF channels $I_{\bar{x}x}$ and $I_{x\bar{x}}$ are described by $\vec{M_\perp}\vec{M_\perp^*}\pm i(\vec{M_\perp}\times\vec{M_\perp^*})_x$ but as the scattering vector lies within the basal plane of the chiral spin structure the chiral term vanishes. The intensity of all three SF-channels are plotted in Fig. \ref{fig:k_vector} c) and d). It can be clearly seen in c) that $I_{\bar{y}y}$ exhibits finite intensity only below $T\approx\SI{5}{\kelvin}$ thus indicating the development of a finite $b$ component of the magnetic structure within the multiferroic phase of \nafegeokomma{,} but exclusively at low temperature. The onset of this component coincides with the lock-in temperature of the incommensurate propagation vector. Fig. \ref{fig:k_vector} d) shows that below the first transition at $T_\text{N}$, the total magnetic scattering signal $I_{\bar{x}x}$ initially features the same amplitude as $I_{\bar{z}z}$, which senses the magnetic moments within the $ac$ plane. This confirms that the magnetic structure is first confined to the $ac$ plane until below $T\approx\SI{5}{\kelvin}$ the discrepancy between $I_{\bar{z}z}$ and $I_{\bar{x}x}$ also indicates the evolving $b$ component of the magnetic structure.

\begin{figure}
 \includegraphics[width=\columnwidth]{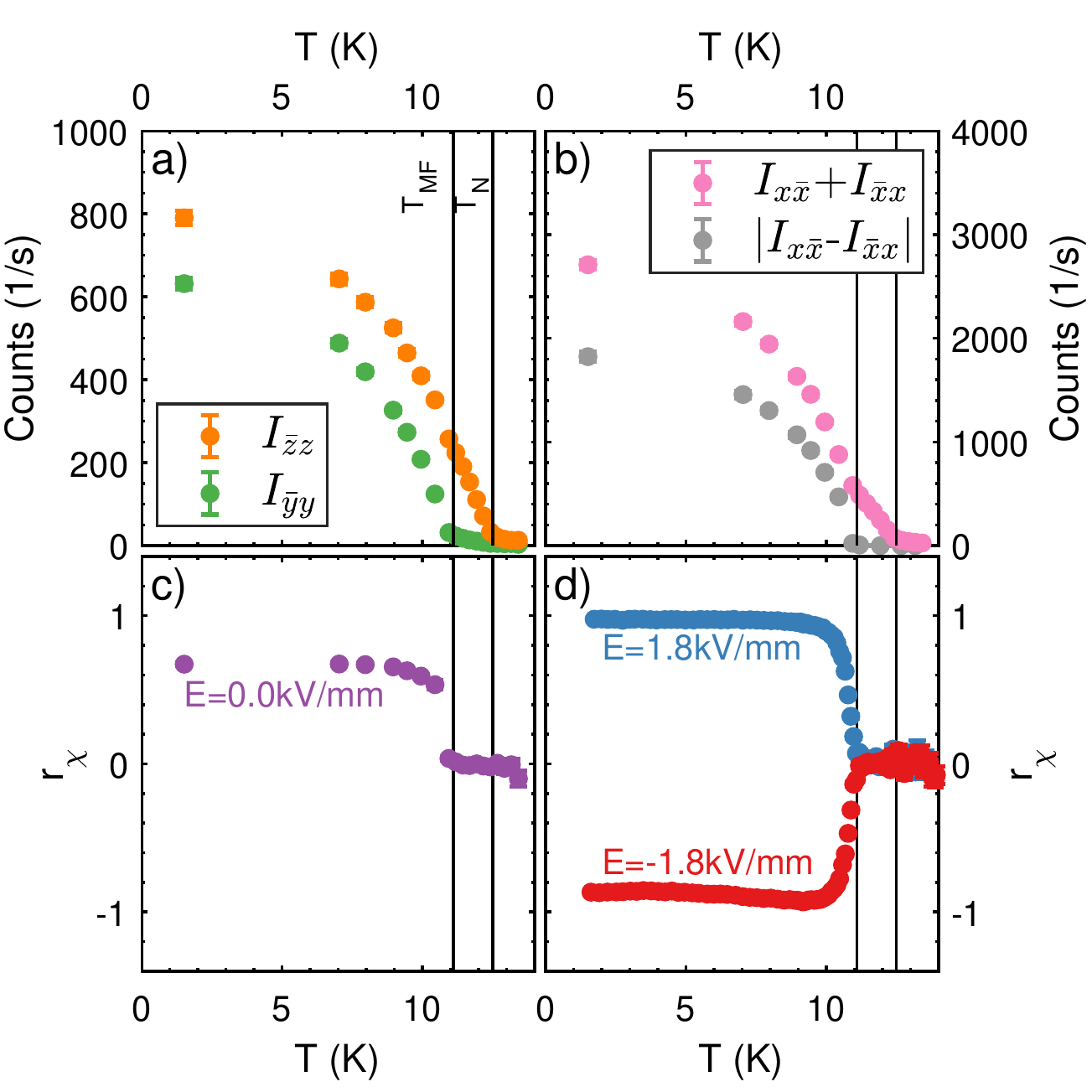}
  \caption{\label{fig:SII_SF_channels}Polarized neutron scattering in (1 0 0)/(0 1 0) geometry. The temperature dependence of SF channels measured at $\vec{Q}=(0.32\:\:3\:\:0.08)$ is shown in a) and b). The $Q$-space position was adjusted with the determined temperature dependent values of the incommensurate propagation vector (see Fig. \ref{fig:k_vector}a) and b)).  In both panels c) and d) the chiral ratio is plotted as a function of temperature and for different applied electric field amplitudes.}
  \label{disp}
 \end{figure}

Sample SII was mounted with scattering geometry (1 0 0)/(0 1 0). In this configuration the SF-channels $I_{\bar{y}y}$ and $I_{\bar{z}z}$ sense the magnetization along $c$ direction and within the $a^*,b$ plane, respectively. The magnetic reflection $\vec{Q}=(0.32\:\:3\:\:0.08)$ is not directly accessible within the given scattering plane but the finite $l$ value can be reached by tilting the goniometer by $\approx\SI{3}{\degree}$, which does not have significant impact on the analysis of respective SF and NSF channels. The temperature dependence of both SF channels is plotted in Fig. \ref{fig:SII_SF_channels} a) and it is visible that the $c$ component of the magnetic structure mainly develops below the second transition $T_\text{MF}$ but nevertheless finite intensity and thus a non-zero component along $c$ is also present in the intermediate phase. The chiral spin structure in the multiferroic phase thus exhibits moments in the $ac$ plane with an additional $b$ component, which develops below $T\approx\SI{5}{\kelvin}$. In the intermediate phase between $T_\text{N}=\SI{12.5}{\kelvin}$ and $T_\text{MF}=\SI{11.1}{\kelvin}$ the SF channel $I_{\bar{z}z}$ possesses extensive intensity compared to $I_{\bar{y}y}$ and as the angle between $\vec{Q}$ and $b$ amounts approximately $\SI{6.9}{\degree}$, the SF channel $I_{\bar{z}z}$ essentially senses the component parallel to $a^*$. With results from both scattering geometries it can be concluded that the SDW structure exhibits moments $\vec{m}$ that are lying essentially in the $a,c$ plane with $\measuredangle(\vec{m},\vec{a^*})\approx\SI{7}{\degree}$.

 \begin{figure}
 \includegraphics[width=\columnwidth]{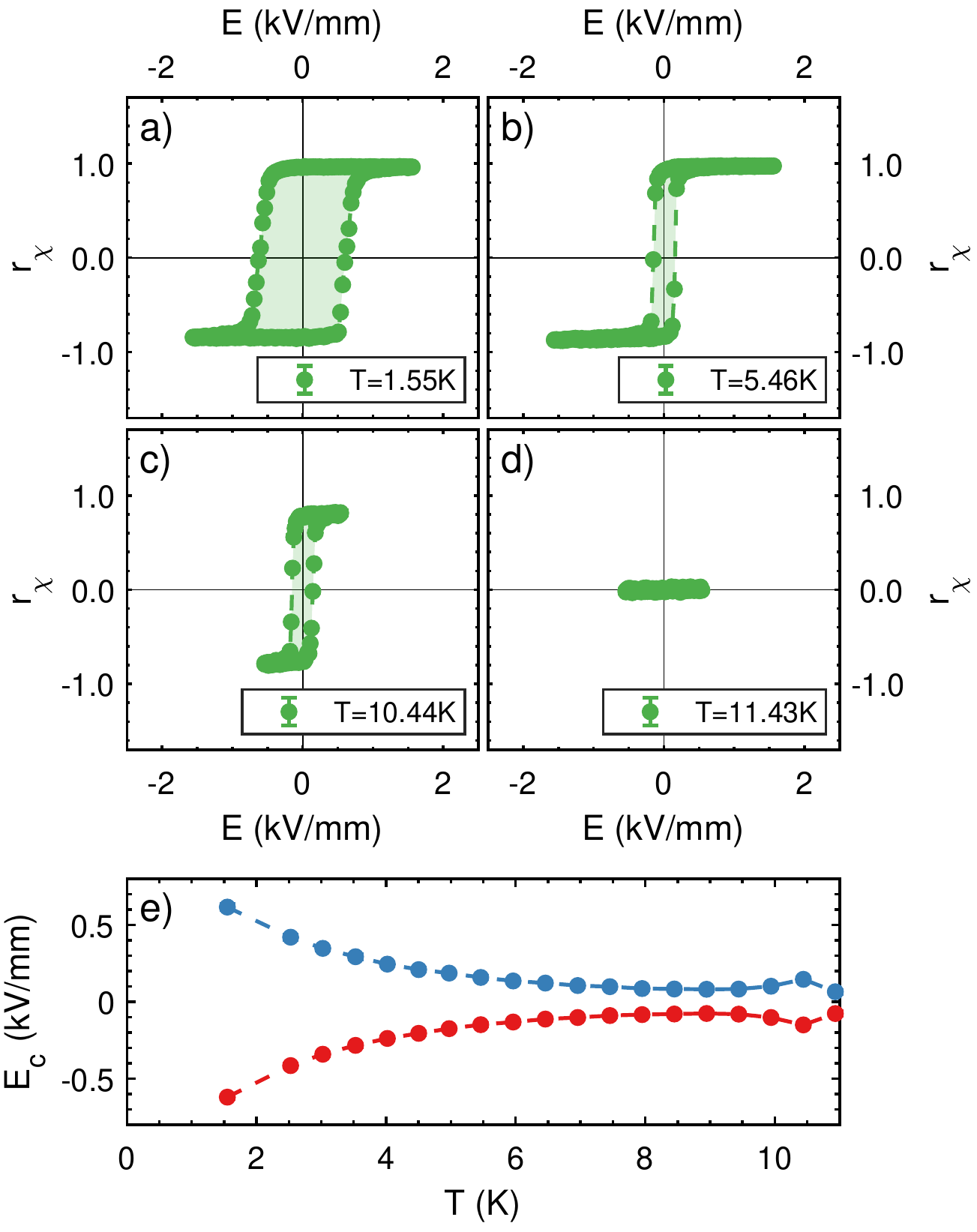}
  \caption{\label{fig:hysteresis}All recorded hysteresis loops of the chiral ratio shown in panels a)-d) were obtained by driving the applied external electric field slowly (with a period duration of about $\SI{45}{\minute}$) between the two polarities. Both slopes of each hysteresis loop were fitted by a hyperbolic tangent in order to estimate the values of respective coercive fields, which are plotted in e). }
  \label{disp}
 \end{figure}
The scattering vector in the scattering configuration of sample SII is almost perpendicular to the basal plane of the chiral structure. Therefore, the chiral term $\pm i(\vec{M_\perp}\times\vec{M_\perp^*})_x$ is finite and contributes to both SF channels $I_{x\bar{x}}$ and $I_{\bar{x}x}$ with a different sign. Hence, the sum and the difference of $I_{x\bar{x}}$ and $I_{\bar{x}x}$ are given by $2\vec{M_\perp}\vec{M_\perp^*}$ and by the absolute value of $-2i(\vec{M_\perp}\times\vec{M_\perp^*})_x$, respectively. Both, the sum and the difference of $I_{x\bar{x}}$ and $I_{\bar{x}x}$ are plotted in Fig. \ref{fig:SII_SF_channels} b) as a function of temperature and it can be seen that the chiral signal develops only below $T_\text{MF}$ within the multiferroic phase. The chiral ratio is defined as $r_\chi=\left(I_{x\bar{x}}-I_{\bar{x}x}\right)/\left(I_{x\bar{x}}+I_{\bar{x}x}\right)$ and for a mono domain sample it corresponds to $\pm i\left(\vec{M_\perp}\times \vec{M_\perp^*}\right)_x/|\vec{M_\perp}|^2$, the ratio of the chiral part with respect to the overall magnetic scattering contribution. The chiral ratio amounts to $\pm1$ only for a completely poled multiferroic state with ideal geometry. However, as the scattering vector $\vec{Q}$ is not perfectly oriented perpendicular to the rotation plane of spins, and due to ellipticity the chiral ratio remains below one. In Fig. \ref{fig:SII_SF_channels} c) the chiral ratio is plotted and it can be seen that even in the absence of an applied electric field, the chiral ratio possesses a finite value within the multiferroic phase and thus indicates an unbalanced domain distribution of multiferroic domains \cite{Finger2010,Biesenkamp2021}. The quasistatic control of the chiral handedness and the multiferroic domain population by external electric fields is discussed in the following section.

 \section{\label{sec:quasistatic_control}Quasistatic control of multiferroic domains}

A static electric field with $|E|=\SI{1.8}{\kilo\volt\per\milli\metre}$ was applied, while cooling the sample. The chiral ratio was measured simultaneously and its temperature dependence is shown in Fig. \ref{fig:SII_SF_channels} d) for both field polarities. It can be seen that the chiral ratio reaches the maximum values of $\approx\pm1$ almost immediately below $T_\text{MF}$,  which indicates a completely poled multiferroic domain distribution and a circular cycloid.
This result illustrates the power of polarized neutron diffraction to solve chiral structures, as only the combination of twice the same magnetic mode can cause a chiral signal in this configuration.
The chiral phase in NaFeG$_2$O$_6$ thus must arise from a single irreducible representation similar to NaFeSi$_2$O$_6$ \cite{Baum2015} but the orientation of the propagation vector differs.
An opposite field polarity entails a reversed chiral ratio and an equivalent temperature dependence documenting the controllability of multiferroic domains with applied electric fields in \nafegeokomma{.} The behavior does not significantly change when lowering the field amplitude to $|E|=\SI{1.2}{\kilo\volt\per\milli\metre}$ and $|E|=\SI{0.6}{\kilo\volt\per\milli\metre}$. In order to quantify the respective values and the temperature dependence of the electric-field amplitude that is needed to reverse the chirality, quasistatic hysteresis loops have been recorded.

Fig. \ref{fig:hysteresis} displays hysteresis loops of the chiral ratio for different temperatures.
The electric field was driven quasistatically between the two polarities corresponding to a period of about 45 minutes for the entire hysteresis loops.
The sign of the vector chirality is invertible down to the lowest temperature at $T=\SI{1.55}{\kelvin}$. While approaching the transition to the intermediate phase, the saturation value of the chiral ratio shrinks slightly and abruptly vanishes, when heating above the multiferroic transition. This was also observed by the cooling sequences with applied static fields (see Fig. \ref{fig:SII_SF_channels} d)). The coercive fields that are needed to invert the sign of the chiral ratio in positive or negative direction were determined by fitting both slopes of the hysteresis loops with a hyperbolic tangent. The obtained coercive fields are plotted in Fig. \ref{fig:hysteresis} e) as a function of temperature. The coercive fields exhibit a symmetric temperature dependence with the smallest values close to the multiferroic transition.

\section{\label{sec:domain_relaxation}Multiferroic domain relaxation}

\begin{figure}
 \includegraphics[width=\columnwidth]{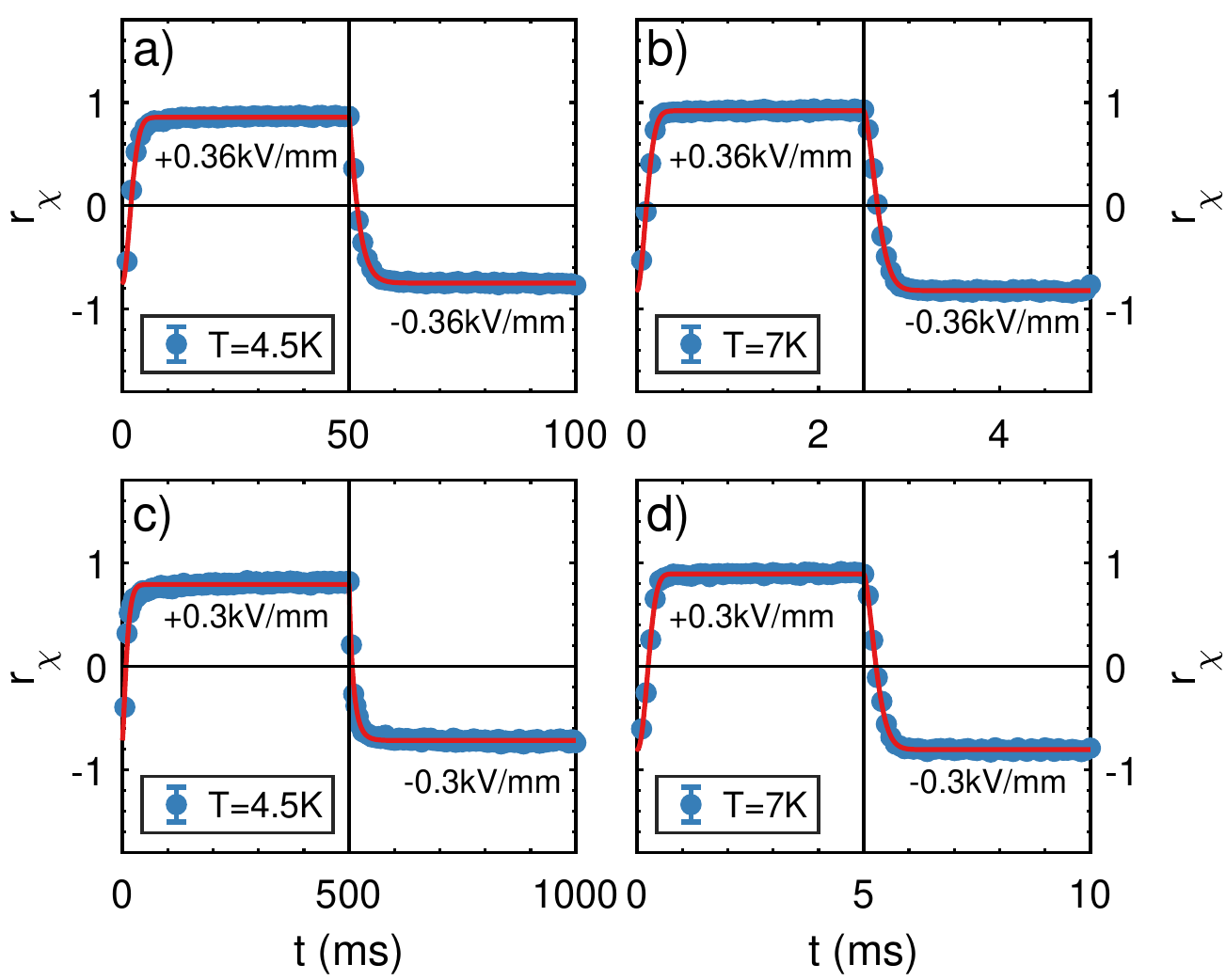}
  \caption{\label{fig:fast_relaxation}All displayed switching curves in a)-d) were recorded by utilizing the stroboscopic method at IN12 and by measuring on sample SII. The blue data points refer to the recorded chiral ratio and the red line corresponds to the fit. The time dependence of the applied electric field follows a rectangular shape with inversion at the beginning and in the middle of each panel.}
  \label{disp}
 \end{figure}

\begin{figure}
 \includegraphics[width=\columnwidth]{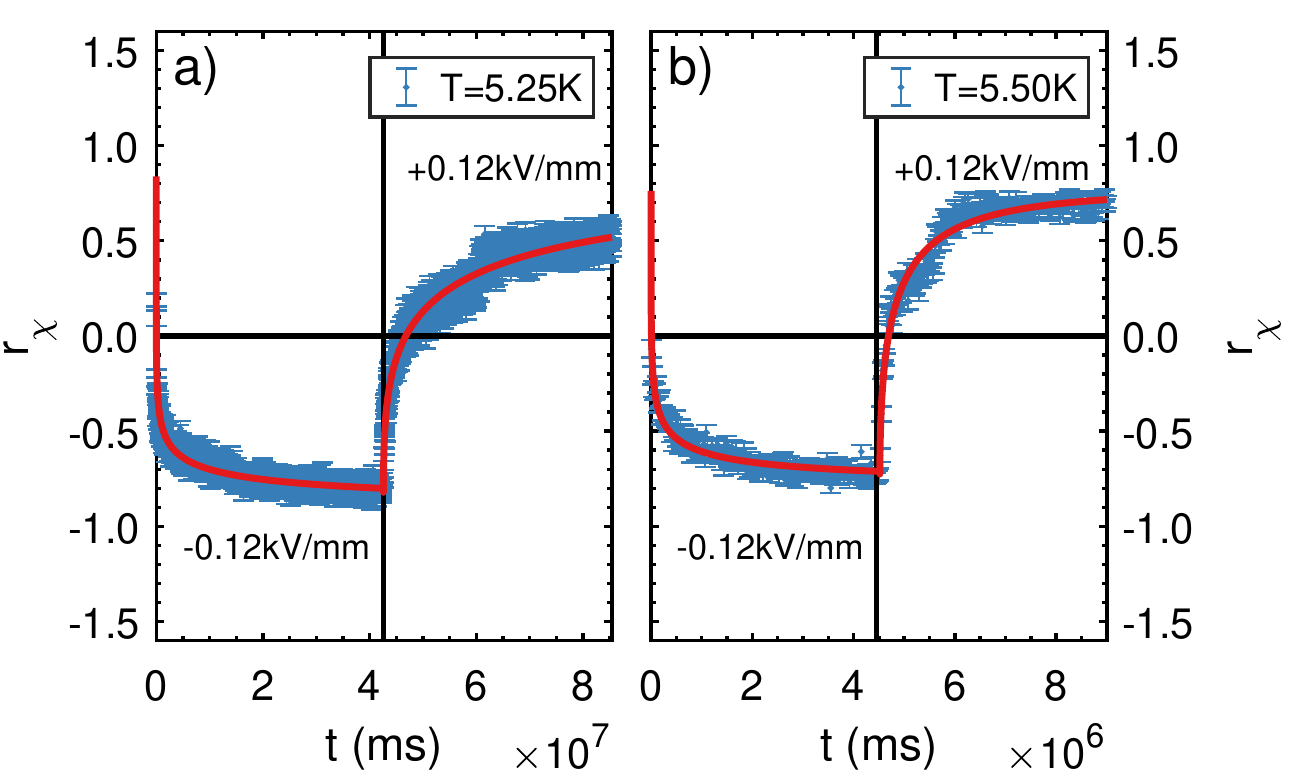}
  \caption{\label{fig:switching_curves_LLB}At 4F1 the relaxation processes of the chiral ratio, which is plotted in a and b), were recorded by counting both SF channels $I_{x\bar{x}}$ and $I_{\bar{x}x}$ as a function of time. The electric field ($|E|=\SI{0.12}{\kilo\volt\per\milli\meter}$) was manually switched after the saturation of the chiral ratio for the respective handedness was reached. For both curves sample SII was measured.}
  \label{disp}
 \end{figure}

In order to investigate the relaxation of multiferroic domains in \nafegeokomma{,} while triggering the  inversion of domains by external electric fields, the stroboscopic setup was installed at IN12 (see section \ref{sec:experimental_methods}). The switching curves of the chiral ratio were recorded for different temperatures and a variety of electric-field amplitudes on sample SII. Fig. \ref{fig:fast_relaxation} displays four exemplary switching curves and it can be seen by comparing the panels horizontally that the relaxation time significantly decreases, when increasing the temperature. Moreover, the vertical arrangement of panels in Fig. \ref{fig:fast_relaxation} displays the field dependence of the relaxation processes. It can be seen that the electric-field strength at fixed temperature also strongly influences the relaxation processes. For all recorded curves, there is no asymmetric relaxation behavior concerning the electric field polarity.

The relaxation process can be described by an exponential relaxation in both directions \cite{stein2020}:
\begin{align}
  \label{eq:switching_a}
r_\chi(t)&=r_a-\left(r_a-r_b\right)\text{exp}\left(-\left(\frac{t}{\tau_a}\right)^{b_1}\right)\\
\label{eq:switching_b}
r_\chi(t)&=r_b-\left(r_b-r_a\right)\text{exp}\left(-\left(\frac{t-t_{1/2}}{\tau_b}\right)^{b_2}\right)
\end{align}
Both parameters $\tau_a$ and $\tau_b$ are the characteristic relaxation times describing the inversion of the chiral ratio from $r_a$ to $r_b$ and from $r_b$ to $r_a$, respectively. Here, $t_{1/2}$ accounts for the reversal of the electric field after half the period. Considering the Ishibashi-Takagi theory, which is based on the Avrami-model, the stretching exponents $b_1$ and $b_2$ describe the dimensionality of the domain wall motion \cite{Ishibashi_1971,Ishibashi_1990,Avrami_1939,Avrami_1940,Avrami_1941}. Both stretching exponents vary between 1 and 2.5 thus indicating low-dimensional domain growth. Approaching $T_\text{N}$ or enhancing the field amplitude result in larger stretching exponents suggesting a higher growth dimensionality or possibly additional nucleation processes. A similar behavior was observed for \tbmno \cite{stein2020}.

The investigation of long relaxation times ($>\SI{1}{\minute}$) does not require the stroboscopic setup, as sufficient statistics are reached during a single switching period. The respective measurements were performed at 4F1. Two switching curves are exemplarily shown in Fig. \ref{fig:switching_curves_LLB} for an electric-field amplitude of $|E|=\SI{0.12}{\kilo\volt\per\milli\meter}$. Although the experiment was performed on the same sample SII, the curves exhibit an asymmetric relaxation behavior contrary to the recorded curves from the IN12 experiment. The system was short circuited at room temperature and during the cooling procedure but nevertheless this asymmetry can be due to residual frozen-in charges or from effective pinning at defects, which provoke a preferred sign of the vector chirality in the multiferroic phase \cite{Finger2010}. Moreover, both curves possess a kink after the second field inversion, which can be related to a time-delayed second nucleation process. However, this effect was resolvable only for the several orders of magnitude slower relaxation processes that were investigated during the 4F1 beamtime.
\begin{figure}
 \includegraphics[width=\columnwidth]{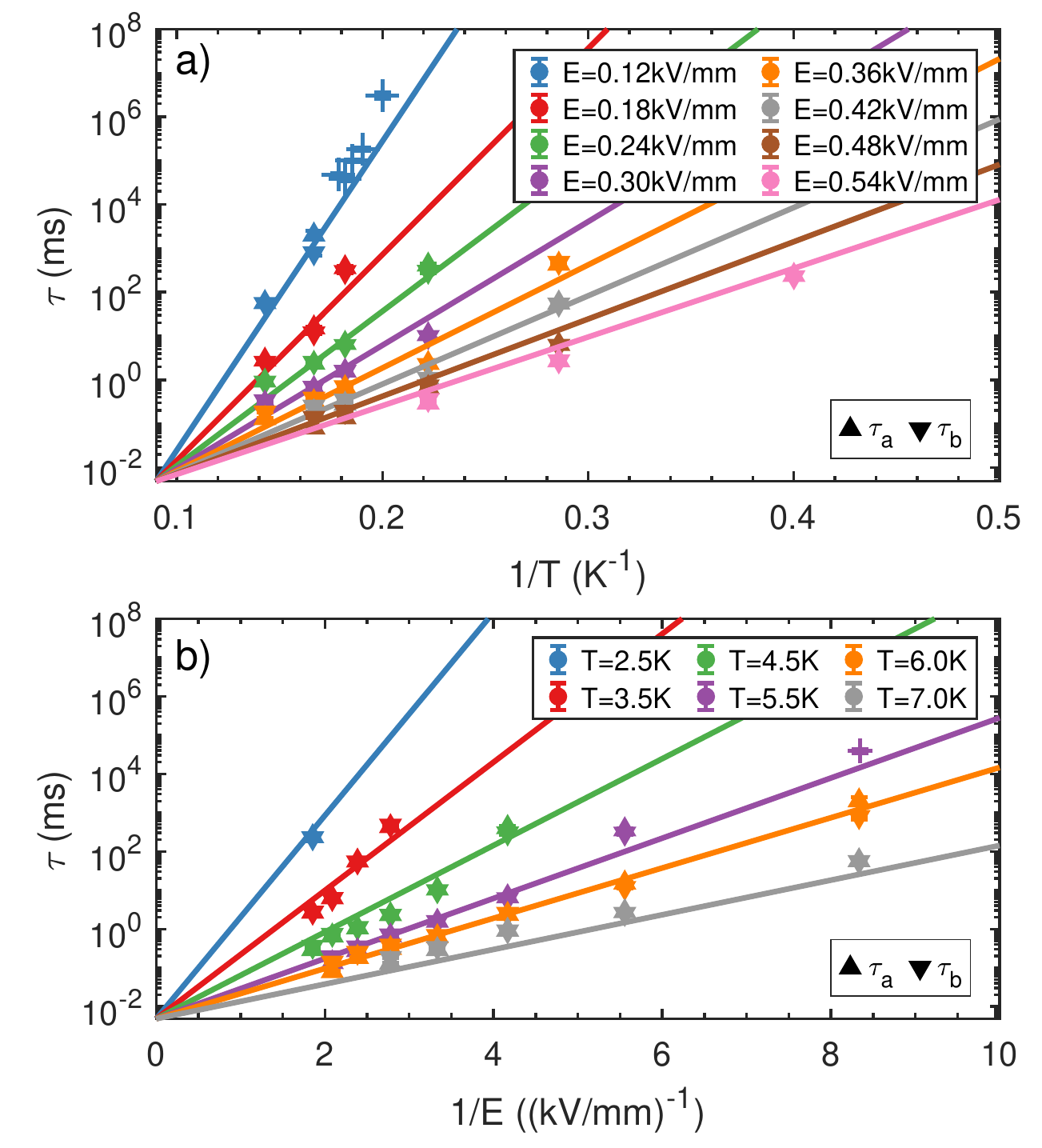}
  \caption{\label{fig:relaxation_times}Panels a) and b) display the temperature and electric-field dependent relaxation times, respectively. Up and down triangles refer to $\tau_a$ and $\tau_b$ from IN12 data, whereas crosses correspond to $\tau_a$ from the recorded 4F1 data. The whole IN12 data set was fitted simultaneously by the combined Arrhenius-Merz law (see equation \ref{eq:arrhenius-merz}) and the plotted colored solid lines refer to the fit results shown for fixed electric fields in a) or fixed temperatures in b). The 4F1 data were not included to the fit.}
  \label{disp}
 \end{figure}
 \begin{figure}
 \includegraphics[width=\columnwidth]{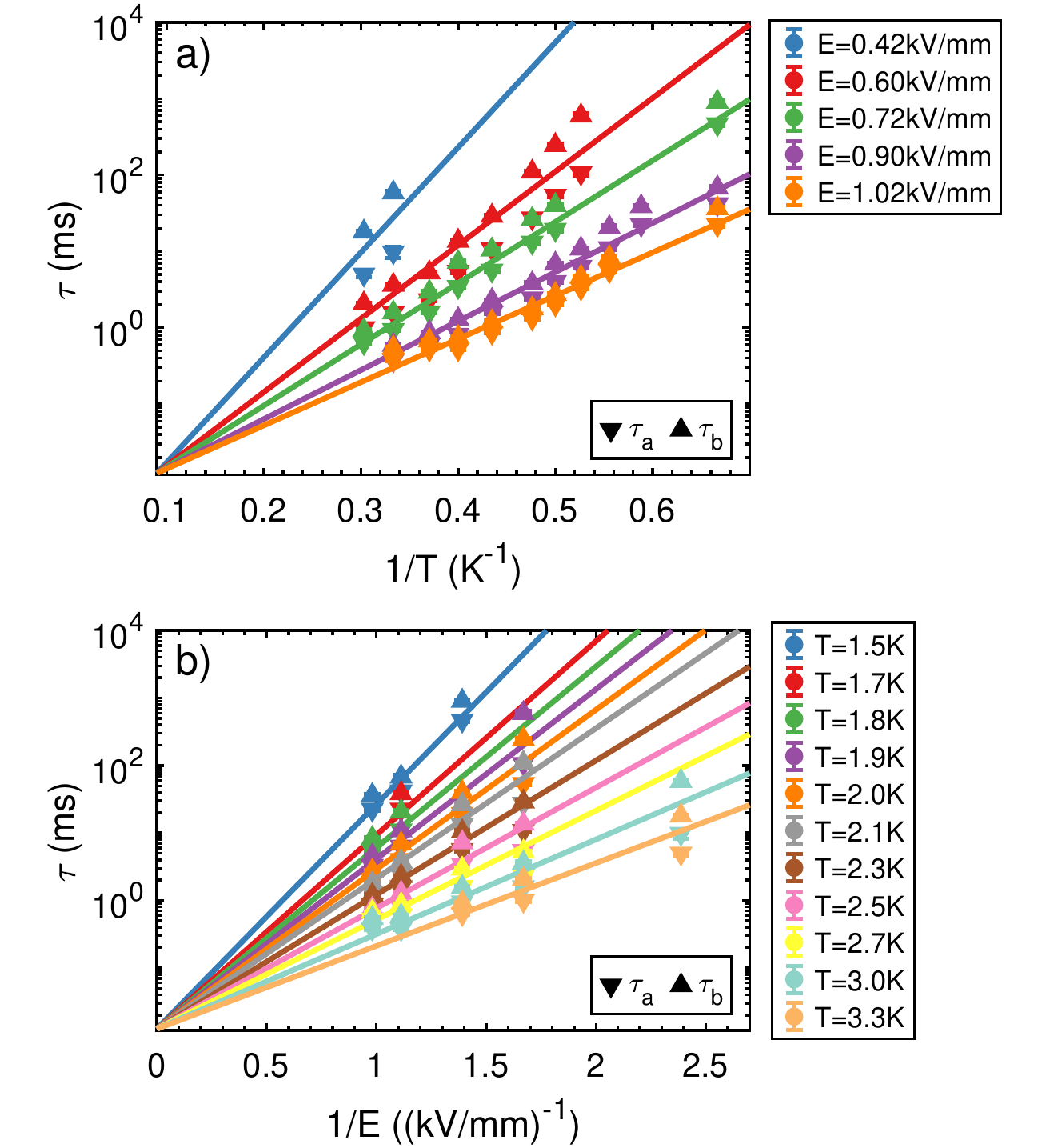}
  \caption{\label{fig:relaxation_times_2}Additional set of measurements of the relaxation times from a second experiment are displayed in panels a) and b). The combined Arrhenius-Merz law is valid even at very low temperatures. A slightly different activation constant and critical relaxation time with respect to results shown in Fig. \ref{fig:relaxation_times} arise from a renewed sample setup (see text).}
  \label{disp}
 \end{figure}

Fig. \ref{fig:relaxation_times} displays the fitted relaxation times $\tau_a$ and $\tau_b$ (see equation \ref{eq:switching_a} and \ref{eq:switching_b}) as a function of inverse temperature and of inverse field with a logarithmic scaling. Up and down arrows mark the fitted relaxation time from IN12 data for switching the electric field to positive and
negative values, respectively. The crosses in a) and b)  mark the fitted relaxation times, which have been obtained from the 4F1 experiment. Combining both experiments at IN12 and 4F1 it is possible to follow the multiferroic relaxation over 7 orders of magnitude in time as a function of temperature and electric field.

The  multiferroic domain relaxation in \tbmno \cite{stein2020} follows a combined Arrhenius-Merz law as a function of temperature and electric-field:
\begin{align}
  \label{eq:arrhenius-merz}
\tau\left(E,T\right)=\tau^*\text{exp}\left(\frac{A_0T_r}{ET}\right)\:\text{with}\:T_r=\frac{T_\text{MF}-T}{T_\text{MF}}
\end{align}
Here $\tau^*$ denotes the fastest possible relaxation time that can be reached at the multiferroic phase transition $T_\text{MF}=T_{\text{N}}$ or at infinitely large fields, and $A_0$ is the activation constant. Thus, $\tau^*$ and $A_0$ are the only two parameters needed to describe the field and temperature dependence of the relaxation.
For \tbmno  the parameters amount to $\tau^*=\SI{0.72}{\milli\second}$ and $A_0=\SI{1483}{\kelvin\kilo\volt\per\mm}$ \cite{stein2020}. This combined Arrhenius-Merz relation describes multiferroic relaxation in the E,T range, where it is determined by classical domain-wall motion. In the studied E,T range, the multiferroic relaxation in TbMnO$_3$ deviates from
the combined Arrhenius-Merz law only when approaching the multiferroic transition, as additional nucleation processes seem to accelerate the domain inversion.

For \nafegeo all relaxation times obtained from IN12 data were simultaneously fitted  with the combined Arrhenius-Merz law. The 4F1 data were excluded from the fit as even a tiny temperature offset between both instruments due to different sample sticks and temperature calibration has significant impact on the fit result. Furthermore, the asymmetry of relaxation and the second nucleation process complicate the precise determination of relaxation times.  Nevertheless, the temperature and electric-field dependent relaxation times $\tau_a$ from 4F1 data are included in Fig. \ref{fig:relaxation_times} (see crosses). Due to the observed asymmetry for long relaxation times only $\tau_a$ is plotted, whereas $\tau_b$ is approximately one order of magnitude larger.

Solid lines in Fig. \ref{fig:relaxation_times} a) and b) correspond to the fit result for different temperatures and different electric-field amplitudes. It is clearly visible that the multiferroic relaxation behavior in \nafegeo is also well described by this combined activation law requiring only two parameters. The obtained critical relaxation time $\tau^*$ and the activation constant $A_0$ amount to $\tau^*=\SI{0.0049(12)}{\milli\second}$ and $A_0=\SI{19.46(88)}{\kelvin\kilo\volt\per\mm}$. During a second experiment at the IN12 spectrometer, the relaxation times at very low temperatures were further studied. Before this beamtime, a tiny piece was cut from sample SII for other measurements and the sample contacts were renewed. With the restored sample setup a slightly different critical relaxation time and activation constant was observed but nevertheless both values ($\tau^*=\SI{0.0125(18)}{\milli\second}$ and $A_0=\SI{13.28(34)}{\kelvin\kilo\volt\per\mm}$) are of the same order of magnitude compared to results from the first experiment. The measured relaxation times as well as the fit are shown in Fig. \ref{fig:relaxation_times_2} and it can be seen that even at low temperature the relaxation follows the activation law indicating that no quantum effects dominate the relaxation behavior, which would lead to a saturation of the relaxation time. For fields between $E=\SI{0.42}{\kilo\volt\per\mm}$ and $E=\SI{1.02}{\kilo\volt\per\mm}$, the domain wall motion in \nafegeo remains thus thermally activated and is not driven by domain wall tunneling down to $T=\SI{1.5}{\kelvin}$. Domain wall tunneling was reported for antiferromagnetic and ferroelectric domain wall motion at low temperature \cite{Kagawa_2016,Brooke_2001}. In \nafegeokomma{,} quantum effects can become relevant at even lower temperature and electric field values.

 \begin{figure}
 \includegraphics[width=\columnwidth]{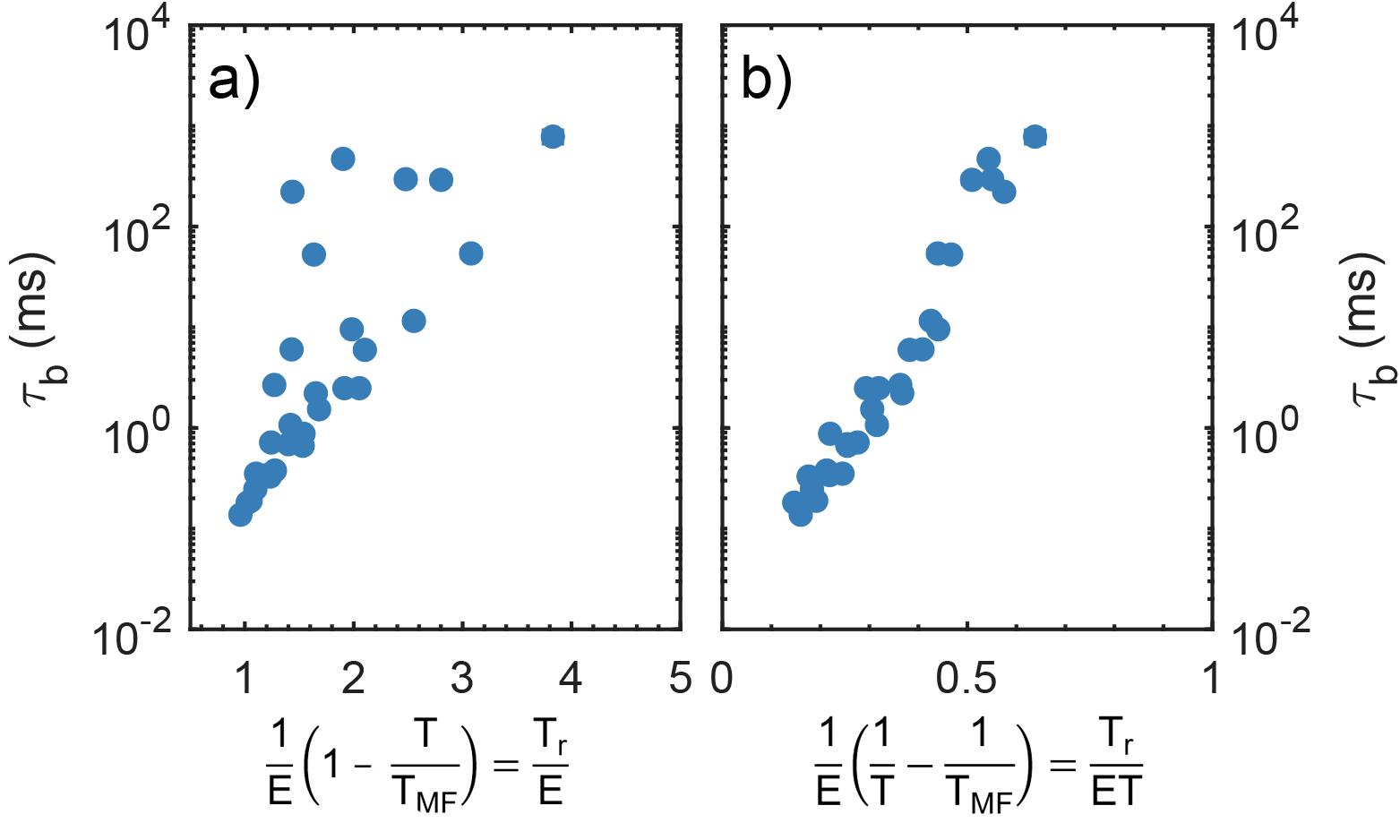}
  \caption{\label{fig:scaling} The different scaling relations for the relaxation times shown in panels a) and b) clearly determine the combined Arrhenius-Merz law to be the correct description of the multiferroic domain inversion. For the comparison only IN12 data from the first experiment (see Fig. \ref{fig:relaxation_times}) were used.}
  \label{disp}
 \end{figure}



Some pure ferroelectrics \cite{Miller_1958,scott2000} exhibit thermally activated relaxation following a scaling law $\tau(E,T)=\tau^*\text{exp}(A_0T_r/E)$, in which
the relaxation only depends on the ratio $T_r/E$.
The measurements on multiferroic \tbmno \cite{stein2020} do not agree with this but follow a scaling with $T_r/(ET)$ as it results from the combined
Arrhenius-Merz law.
Fig. \ref{fig:scaling} compares both scaling laws for \nafegeo  and clearly shows that the combined Arrhenius-Merz law is the appropriate description also
for this material.

The relaxation of multiferroic domains differs from the relaxation behavior in conventional ferroelectrics \cite{Miller_1958,scott2000,stein2020}. While in a ferroelectric material the nucleation process, the forward growth and the subsequent sideways expansion of domains can equally contribute to the global domain relaxation process and thus render its description complicated \cite{Miller_1958,scott2000}, multiferroic domain inversion is dominantly determined by the sideways growth of domains \cite{stein2020,Hoffmann2011a,Meier2009,Matsubara2015}.
Therefore, thermally activated domain wall motion and a combined Arrhenius-Merz law \cite{stein2020}
describe the multiferroic domain inversion over a wide range of temperatures and fields, spanning over 7 orders of magnitude in time.
However, close to the multiferroic transition and for very high fields additional nucleation processes will accelerate the relaxation, and at very low temperatures and
electric fields quantum tunneling can become relevant.

 \section{\label{sec:spin_wave}Spin-wave velocity}
  \begin{figure}
 \includegraphics[width=\columnwidth]{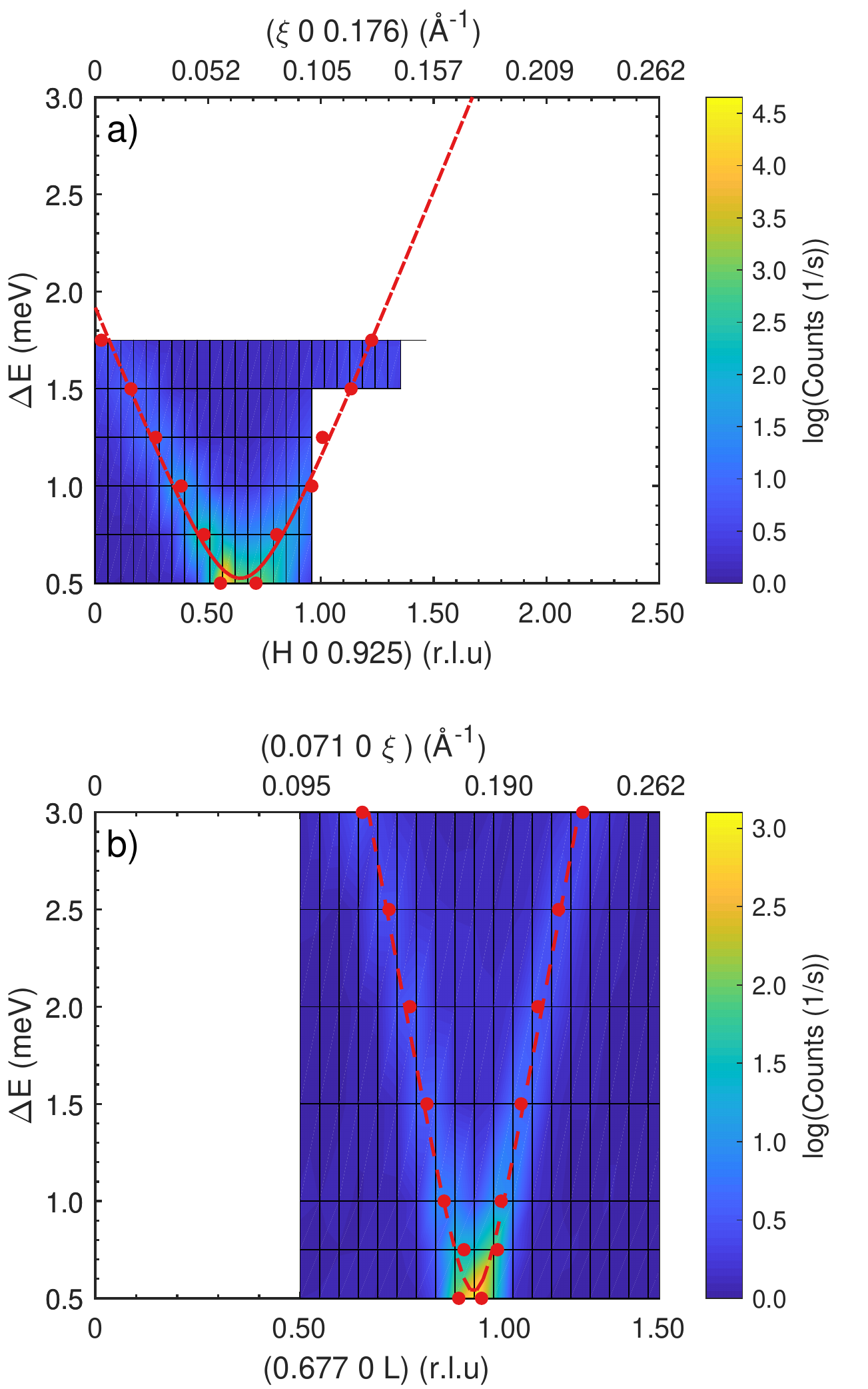}
  \caption{\label{fig:spin_wave}The plots in a) and b) display the recorded constant-energy scans at $T=\SI{3.9}{\kelvin}$ around the incommensurate zone center $\vec{Q}=(0.677\:\:0\:\:0.9252)$ along $H$ and $L$, respectively. Red data points refer to the fitted peak maxima and the red lines correspond to fits of the respective dispersion. }
  \label{disp}
 \end{figure}

Since the spin-wave velocity can limit the multiferroic domain wall motion \cite{stein2020,Hoffmann2011a} we analyzed the dispersion of acoustic magnon branches at the 4F2 triple-axis spectrometer. The respective measurements were carried out on sample SI, which was mounted with scattering geometry (1 0 0)/(0 0 1). For $\Delta E\geq\SI{0.5}{\milli\electronvolt}$, constant-energy scans were measured around the incommensurate zone center $\vec{Q}=(0.677\:\:0\:0.925)$ along $H$ and $L$ directions at $T=\SI{3.9}{\kelvin}$. Below $\Delta E=\SI{0.5}{\milli\electronvolt}$ it was not possible to separate the excitation from the elastic line due to the finite resolution.

Fig. \ref{fig:spin_wave} maps the measured scattering along both recorded directions. All constant-energy scans were fitted with two Gaussian functions and the obtained peak-center positions of excitations are marked by red data points in a) and b). The dispersion can be approximately described by $\omega(q)=\sqrt{(\Delta^2+v_g^2q^2)}$, with $v_g$ the group velocity of the spin wave and $\Delta$ the gap energy. The fits for respective directions (red dashed lines in Fig. \ref{fig:spin_wave}) yield $v_g=\SI{662(19)}{\meter\per\second}$ along $H$, $v_g=\SI{1460(36)}{\meter\per\second}$ along $L$ and a spin gap of approximately $\SI{0.53}{\milli\electronvolt}$ for both directions. The magnon dispersion is thus anisotropic with a steeper slope along the direction of zigzag chains, but this anisotropy is moderate and the material should not be considered as a low-dimensional system. The ferroelectric polarization is predominantly aligned along the $a$ direction \cite{Ackermann_2015} and for all respective experiments on the multiferroic domain inversion, the electric field was applied parallel to it. In multiferroics the initial reverted domains extend along the direction of ferroelectric polarization before the subsequent sideways growth dominates the overall relaxation process \cite{stein2020,Hoffmann2011a}. Considering the determined spin-wave velocities, a single needle-shaped domain in the middle and the geometry of sample SII (for which the relaxation behavior was measured), the fastest possible domain relaxation is of the order of $\SI{e-3}{\milli\second}$, which agrees with the observed critical relaxation time $\tau^*$ (see section \ref{sec:domain_relaxation}). However, a striped pattern of more than one needle-shaped domain as well as nucleation of additional germs will further reduce the multiferroic relaxation time, because both effects reduce the distances to be overcome by the domain wall motion.

 \section{\label{sec:level1}Conclusion}

We report investigations on the magnetic structure as well as on the multiferroic domain relaxation in \nafegeo utilizing neutron scattering experiments on single crystals. With neutron polarization analysis it is possible to separate the magnetic components of the long-range order in both phases. Thereby we can  confirm the proposed incommensurate spin-density wave with moments along the $a$ direction for the intermediate phase and a chiral spin structure with components along $a$ and $c$ below the multiferroic transition. In addition there is a locking-in of the incommensurate propagation vector at $T\approx\SI{5}{\kelvin}$ accompanied with the emergence of a $b$-component, whose presence was so far controversially discussed \cite{Drokina_2011,Redhammer_2011}.

The multiferroic domain inversion was investigated by quasi-static as well as by time-resolved measurements of the vector chirality. Poling sequences and hysteresis loops confirm the invertibility of its sign and show low coercive fields as a function of temperature. As reported for \tbmno \cite{stein2020},  time-resolved measurements of the rapidly switched chiral ratio reveal the combined Arrhenius-Merz law to describe the relaxation behavior in \nafegeo over at least 7 orders of magnitude in time. Even at low temperatures of the order of 1.5\,K the multiferroic domain inversion is well described by thermally activated domain-wall motion.

Since \nafegeo allows one to study the multiferroic domain relaxation over a large temperature range (down to $T/T_\text{MF}\approx 0.14$), we can analyze the scaling of electric field and reduced temperature $T_\text{r}$. While a simple scaling $f\left(\frac{T_r}{E}\right)$ can be applied to ferroelectrics \cite{Miller_1958,scott2000} such scaling can be excluded for the multiferroic domain relaxation in \nafegeokomma{,} which follows the $f\left(\frac{T_r}{TE}\right)$ scaling predicted by the Arrhenius-Merz law.


In conclusion, \nafegeo exhibits similar to \tbmno a simple field and temperature dependence of multiferroic domain inversion that is well described by thermally activated domain-wall motion with an activation field given by $A_0\frac{T_\text{r}}{T}$.

The neutron scattering data from the IN12 diffractometer are available \cite{dataIN12}.

\section{\label{sec:level1}Acknowledgements}

This work was funded by the Deutsche Forschungsgemeinschaft (DFG,
German Research Foundation) - Project number 277146847 - CRC 1238, projects A02, B04 and by
the Bundesministerium f\"ur Bildung und
Forschung - Project number 05K19PK1. This work is based upon experiments performed at the IN12 instrument operated by JCNS at the Institute Laue-Langevin (ILL), Grenoble, France. The authors gratefully acknowledge the financial support provided by JCNS to perform the neutron scattering measurements at the at the Institute Laue-Langevin (ILL), Grenoble, France.

\nocite{apsrev41Control}
\bibliographystyle{apsrev4-1}

\end{document}